# On the Square Speed of Sound in High-Energy Collisions: Range of Values and How to Understand It

Ting-Ting Duan[1,a)], Fu-Hu Liu[1,b)], Khusniddin K. Olimov[2,3,c)]

[1]*Institute of Theoretical Physics, State Key Laboratory of Quantum Optics Technologies and Devices & Collaborative Innovation Center of Extreme Optics, Shanxi University, Taiyuan 030006, China*

[2]*Laboratory of High Energy Physics, Physical-Technical Institute of Uzbekistan Academy of Sciences, Chingiz Aytmatov Str. 2b, Tashkent 100084, Uzbekistan*

[3]*Department of Natural Sciences, National University of Science and Technology MISIS (NUST MISIS), Almalyk Branch, Almalyk 110105, Uzbekistan*

**Abstract:** After reviewing the sound speeds in various forms and conditions of matter, we investigate the sound speed of hadronic matter that has decoupled from the hot and dense system formed during high-energy collisions. We comprehensively consider factors such as energy loss of the incident beam, rapidity shift of leading nucleons, and the Landau hydrodynamic model for hadron production. The sound speed is related to the width or standard deviation of the Gaussian rapidity distribution of hadrons. The extracted square speed of sound lies within a range from 0 to 1/3 in most cases. For scenarios exceeding this limit, we also provide an explanation.

**Keywords:** square speed of sound, energy loss, rapidity shift, Gaussian rapidity distribution

**PACS numbers:** 12.40.Ee, 13.85.Ni, 13.87.Ce

## 1. Introduction

In high-energy collisions, the hot and dense system generated in regions with violently compressed participant nucleons exhibits a wealth of characteristics. Quantities such as pressure, energy density, entropy density, temperature, volume, and others can be measured or extracted using appropriate methods. In addition to the equation of state, which defines the ratio between pressure and energy density, both sound speed ($c_s$) and square speed of sound ($c_s^2$) are important extractable quantities that contribute to our understanding of collision system evolution processes as well as accelerated expansion phenomena in cosmology. For hadronic matter resulting from decoupling within a hot and dense environment, $c_s^2$ can be determined by taking the partial derivative of pressure with respect to energy density at constant entropy density; alternative formulations may arise due to relationships among thermodynamic quantities [1].

a) 202312602001@email.sxu.edu.cn

b) Correspondence: fuhuliu@163.com; fuhuliu@sxu.edu.cn, ORCID ID 0000-0002-2261-6899

c) Correspondence: khkolimov@gmail.com; kh.olimov@uzsci.net, ORCID ID 0000-0002-1879-8458

The values for $c_s^2$ vary across different models and types of matter while maintaining similar definitions. Beyond just high-energy collision systems characterized by their hot and dense state, other forms and conditions—such as neutron stars [2, 3], dark matter [4], dark energy [5–7], magnetized nuclear matter [8], gas composed of closed strings [9], and adiabatic evolution in heavy fields [10]—exhibit significant differences in their respective ranges for $c_s^2$. Qualitative discussions on the various ranges of $c_s^2$ in different forms and conditions of matter are essential for enhancing our understanding of the ranges of $c_s^2$ associated with hadronic matter decoupled from hot and dense system.

Within neutron stars, a quantitative Bayes factor analysis indicates that the conformal bound $0 \leq c_s^2 \leq 1/3$ is violated [3], despite applying the same definition of $c_s^2$ as used for hadronic matter. As energy density increases, $c_s^2$ initially rises before exhibiting fluctuations—either large or small [2, 3]. The range of $c_s^2$ for neutron stars is found to be $0 \leq c_s^2 \leq 1$. For dark matter sound speed constrained by galactic scales, cases involving modified and extended Chaplygin gas yield values such that $0 \leq c_s^2 < 10^{-8}$ [4]. This value is indeed very small and presents a more competitive constraint compared to previous limits derived from anisotropic cosmic microwave background measurements and baryonic acoustic oscillations which exhibit power spectra.

Using a constant linear equation of state, the behavior of dark energy perturbations during radiation and matter epochs in both synchronous and conformal Newtonian gauges has been analytically studied in reference [5]. It was demonstrated that the range for sound speed can be constrained to $10^{-4} \leq c_s^2 \leq 1$, which contradicts results obtained from dark matter studies. Employing the Markov chain Monte Carlo method, a global analysis based on current observational data from anisotropic cosmic microwave background suggests that constraints on $c_s^2$ remain quite weak. Furthermore, if sound speed becomes negative, the system exhibits instability due to divergent classical perturbations [6]. Predictions across all scalar field models featuring standard kinetic terms indicate that $c_s^2 = 1$. However, models such as $k$-essence predict an effective sound speed significantly greater than one. Although it remains challenging to quantitatively measure the value of $c_s^2$ associated with dark energy directly, it can influence measurements related to other parameters [7].

Within the framework of the nonlinear Walecka model, a calculation for nuclear matter subjected to a background magnetic field at finite temperature and baryon chemical potential reveals that the presence of a magnetic field induces anisotropy in $c_s^2$. Consequently, the parallel (or longitudinal) and perpendicular (or transverse) components, denoted as $c_s^{2(\parallel)}$ and $c_s^{2(\perp)}$, respectively, can exhibit variation [8]. Furthermore, related calculations indicate that $0 < c_s^2 < 1$ in nuclear matter at high baryon chemical potential, irrespective of whether a magnetic field is present or absent [8]. According to reference [9], matter in string gas cosmology can be conceptualized as a gas of closed strings, which suggests that $0 < c_s^2 \ll 1$. In accordance with the framework of adiabatic evolution in heavy fields [10], $c_s^2$ can be expressed in terms of the effective mass of heavy fields and the turning rate of the background trajectory within multi-field space. This leads to a suppressed sound speed characterized by $0 < c_s^2 \ll 1$.

From this introduction—which provides only select examples regarding sound speed—it is evident that values for $c_s^2$ span a wide range. Even within nuclear matter, one finds that $0 < c_s^2 < 1$. It is therefore intriguing to investigate $c_s^2$ further and constrain its range during high-energy collisions. In this paper, we study $c_s^2$ under conditions prevalent in high-energy collisions based on energy loss from incident beams, rapidity shifts observed in leading nucleons, and utilizing Landau hydrodynamic model for hadron production. We elucidate its significance according to its defined ranges.

The remainder of this mini-review article is organized as follows: Section 2 gives formula description of square speed of sound; Section 3 presents possible range of square speed of sound and related explanation; finally, Section 4 provides a summary.

## 2. Formula description of square speed of sound

Let us denote by $E_{T\mathrm{beam}}$, $E_{P\mathrm{beam}}$, and $m_N$ the energy of the target beam, energy of the projectile beam, and mass of protons, respectively. During collisions, we assume an energy loss rate denoted by $k$ for either incident target or projectile nucleons. Following these collisions, we can express both energy ($E_{\mathrm{LT}}$) and momentum ($p_{\mathrm{LT}}$) for leading target nucleons alongside their counterparts—energy ($E_{\mathrm{LP}}$) and momentum ($p_{\mathrm{LP}}$)—for leading projectile nucleons through

$$E_{\mathrm{LT}} = (1-k)E_{T\mathrm{beam}}, \tag{1}$$

$$p_{\mathrm{LT}} = \sqrt{E_{\mathrm{LT}}^2 - m_N^2}, \tag{2}$$

$$E_{\mathrm{LP}} = (1-k)E_{P\mathrm{beam}}, \tag{3}$$

and

$$p_{\mathrm{LP}} = \sqrt{E_{\mathrm{LP}}^2 - m_N^2}, \tag{4}$$

respectively. For the same size of target and projectile, one has

$$E_{T\mathrm{beam}} = E_{P\mathrm{beam}} = \frac{\sqrt{s_{NN}}}{2}, \tag{5}$$

where $\sqrt{s_{NN}}$ denotes the center-of-mass energy per nucleon pair.

For the leading nucleon, its transverse momentum is nearly zero, while its longitudinal momentum is approximately equal to its total momentum. In comparison with mid-rapidity, typically defined by $y_C = 0$, the rapidity shifts of the leading target and projectile nucleons are denoted as

$$y_{\mathrm{LT}} \approx -\frac{1}{2}\ln\left(\frac{E_{\mathrm{LT}} + p_{\mathrm{LT}}}{E_{\mathrm{LT}} - p_{\mathrm{LT}}}\right) \tag{6}$$

and

$$y_{LP} \approx \frac{1}{2}\ln\left(\frac{E_{\rm LP}+p_{\rm LP}}{E_{\rm LP}-p_{\rm LP}}\right) \tag{7}$$

respectively. The total rapidity shift from the leading target nucleon to leading projectile nucleon can be given by

$$L = y_{\rm LP} - y_{\rm LT} \approx \frac{1}{2}\ln\left[\left(\frac{E_{\rm LT}+p_{\rm LT}}{E_{\rm LT}-p_{\rm LT}}\right)\left(\frac{E_{\rm LP}+p_{\rm LP}}{E_{\rm LP}-p_{\rm LP}}\right)\right]. \tag{8}$$

The total rapidity shift represents the distribution range of particles in one-dimensional rapidity $y$ space. According to the Landau hydrodynamic model for hadron production in high-energy collisions [11–14], $y$ follows an approximate Gaussian distribution characterized by a width or standard deviation $\sigma$ [14, 15]. Based on properties of Gaussian distributions, it can be expected that approximately 99.7% of particles will fall within the rapidity range of $[-3\sigma, 3\sigma]$. If one considers a range of $6\sigma$ as representing the total rapidity shift $L$, then one has

$$\sigma \approx \frac{1}{6}L. \tag{9}$$

The Landau hydrodynamic model for hadron production results in the relationship between $\sigma$ and $c_s^2$ to be [11, 16–21]

$$\sigma = \sqrt{\frac{8}{3}\frac{c_s^2}{1-c_s^4}\ln\left(\frac{\sqrt{s_{NN}}}{2m_N}\right)}. \tag{10}$$

Then, the square speed of sound is expressed by

$$c_s^2 = \frac{1}{3\sigma^2}\left[-4\ln\left(\frac{\sqrt{s_{NN}}}{2m_N}\right) + \sqrt{16\ln^2\left(\frac{\sqrt{s_{NN}}}{2m_N}\right) + 9\sigma^4}\right] \tag{11}$$

or

$$c_s^2 = -\frac{4}{3\sigma^2}\ln\left(\frac{\sqrt{s_{NN}}}{2m_N}\right) + \sqrt{\left[\frac{4}{3\sigma^2}\ln\left(\frac{\sqrt{s_{NN}}}{2m_N}\right)\right]^2 + 1}. \tag{12}$$

It is important to note that an analytical solution for non-conformal and viscous Landau hydrodynamics has been presented, along with a proposed form of the speed of sound in ref. [22]. This proposed solution can be generalized to ideal and conformal cases when appropriate limits are applied. Furthermore, this solution successfully describes the rapidity spectra observed in existing experimental data and offers a novel approach for probing the equation of state as well as extracting the speed of sound during heavy ion collisions. In ref. [22], another general relationship between $\sigma^2$ and $c_s^2$ can also be found. When compared to conformal solutions [11, 16–21], the non-conformal solution exhibits a slightly lower peak and a marginally larger $\sigma^2$ in rapidity distribution.

## 3. Possible range of square speed of sound and related explanation

Building upon these discussions, one can conveniently study the relationships between $L$ and $\sqrt{s_{NN}}$, as well as between $c_s^2$ and $\sqrt{s_{NN}}$. To determine $L$, value for parameter $k$ is required. In proton-proton collisions, the binary (nucleon-nucleon) collision number ($\nu$) naturally equals 1. However, in proton-nucleus and nucleus-nucleus collisions, the maximum binary collision number ($\nu_{\max}$) associated with a specific incident nucleon significantly exceeds 1. For instance, in proton-aluminum, proton-copper, and proton-uranium collisions where events exhibit maximum binary collision numbers of 9, 12, and 16 respectively; these correspond to fractions in their respective $\nu$ distributions being approximately 0.004%, 0.002%, and 0.010% [23]. Conversely, for these three specific types of collisions at $\nu = 1$, corresponding fractions are found to be about 47.177%, 34.479%, and 23.000%. It is evident that from $\nu = 1$ to $\nu = \nu_{\max}$, there is a significant decrease in corresponding fractions.

Some investigations indicate that, in multiple binary collisions, at least one instance experiences an energy loss of 50%, while the remaining instances incur an energy loss ranging from 10% to 25% [23–26]. To conduct a comprehensive study, it is advisable to consider as many values as possible within the range of $0.5 < k < 1$. The dependence of $L$ on $\sqrt{s_{NN}}$ is illustrated in Figure 1. For comparison purposes, we set $k$ values at 0.5, 0.6, 0.7, 0.8, and 0.9; these correspond to various curves marked with different colors in the panel. It is evident that $L$ increases significantly with rising $\sqrt{s_{NN}}$ for all selected $k$ values. Additionally, in the initial region of $\sqrt{s_{NN}}$, $L$ exhibits a linear increase with respect to $\ln\sqrt{s_{NN}}$. Naturally, smaller $k$ values result in larger $L$.

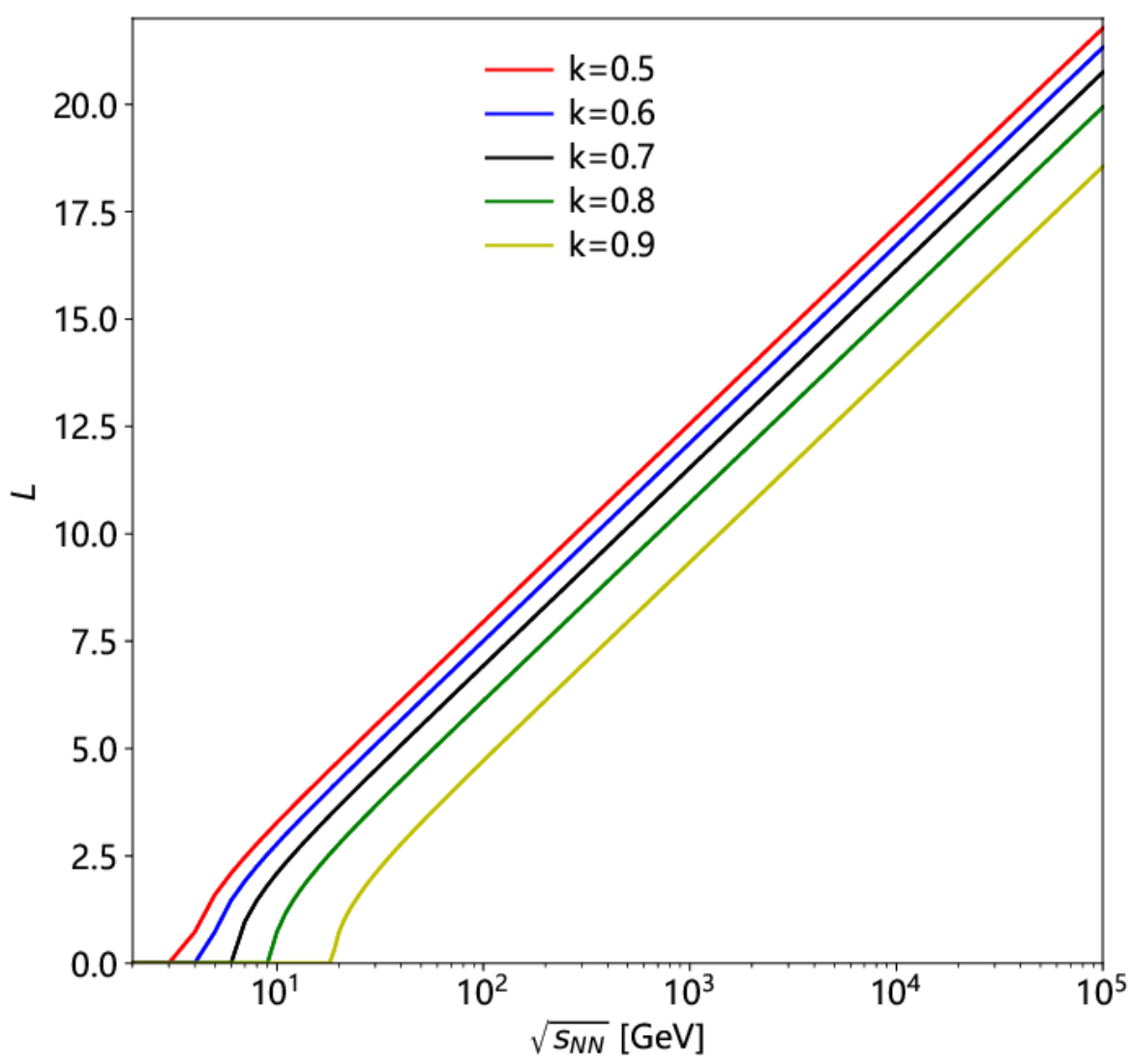

Fig. 1. Dependence of $L$ on $\sqrt{s_{NN}}$, in which $k = 0.5$, 0.6, 0.7, 0.8, and 0.9 which corresponds to various curves with different colors marked in the panel.

Figure 2 presents the relationship between $c_s^2$ and $\sqrt{s_{NN}}$ at $\sigma = L/6$, which reflects the required and reasonable correlation between $\sigma$ and $L$. For comparative analysis, we again utilize $k$ values of 0.5, 0.6, 0.7, 0.8, and 0.9 corresponding to distinct curves represented by different colors within the panel. The results demonstrate that $c_s^2$ increases markedly with increasing $\sqrt{s_{NN}}$ across various $k$ values; however, in some cases—specifically above hundreds of GeV—the growth rate of $c_s^2$ begins to gradually decelerate, where smaller $k$ leadings to larger $c_s^2$ outcomes; notably many calculated $c_s^2$ remain below 1/3. Only when $\sqrt{s_{NN}}$ exceeds approximately ten TeV do some instances yield $c_s^2$ greater than 1/3.

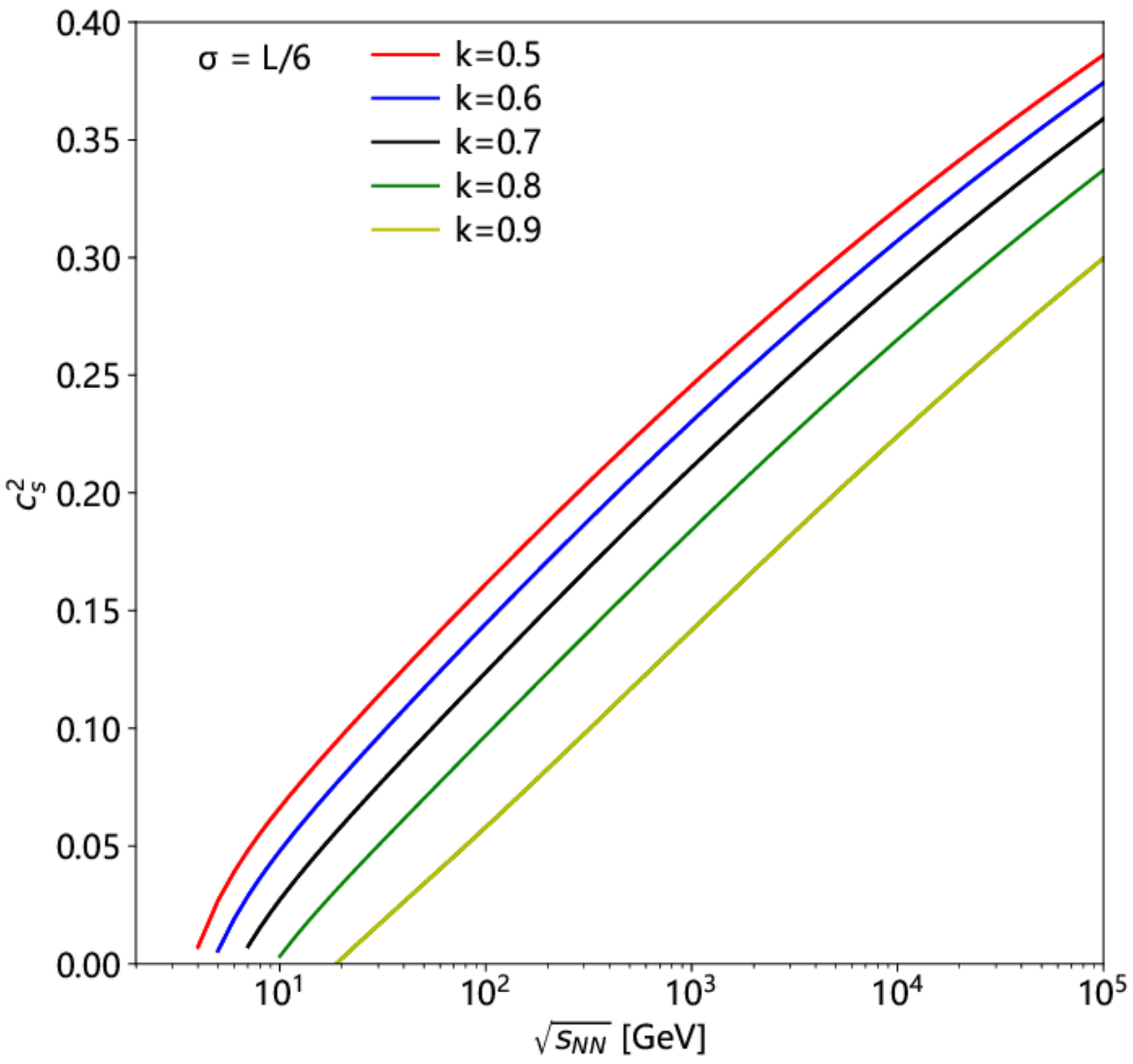

Fig. 2. Dependence of $c_s^2$ on $\sqrt{s_{NN}}$ at $\sigma = L/6$, in which $k = 0.5, 0.6, 0.7, 0.8$, and 0.9 which corresponds to various curves with different colors marked in the panel.

The dependence of $c_s^2$ on $\sqrt{s_{NN}}$ is illustrated in Figure 3 for different values of $\sigma$, specifically $\sigma = 0.2L$ (a), $0.4L$ (b), $0.6L$ (c), and $0.8L$ (d). In each panel, the parameter $k = 0.5, 0.6, 0.7, 0.8$, and 0.9, corresponding to various curves represented by different colors, is employed. Notably, all four values of $\sigma$ used in Figure 3 are greater than the value of $\sigma = L/6$ utilized in Figure 2; however, it should be noted that $\sigma = 0.2L$ is closest to this reference point. In all cases examined, there is a significant increase in $c_s^2$ with rising values of $\sqrt{s_{NN}}$. Furthermore, at energies exceeding hundreds of GeV—even down to approximately ten GeV—the growth rate of $c_s^2$ begins to decelerate gradually. It is evident that while $c_s^2 > 1/3$ holds true for most scenarios considered here, it remains less than one throughout the analysis presented in Figure 3; however, it must be emphasized that the values depicted are not physically realizable under standard conditions but serve primarily as a means to explore broader trends when parameters exceed physical limits.

Although the relationships between $\sigma$ and $L$ shown in Figure 3 may lack physical justification, the mathematical dependence of $c_s^2$ on $\sqrt{s_{NN}}$ can still be derived analytically. This study proposes a reasonable relationship: namely, $\sigma = L/6$, which establishes a connection between Gaussian distribution width $\sigma$ and rapidity shift $L$. If rapidity does not conform to Gaussian distribution at very high energies (e.g., above several hundred GeV), yet adheres instead to a superposition model comprising two or three Gaussian distributions, then one would identify two or three emission sources along with their respective widths $\sigma$ and square speed of sound $c_s^2$. Although each $c_s^2$ can be obtained from the relationship with $\sigma$ [Eqs. (10)-(12)], each $\sigma$ can no longer be obtained from the relationship with $L$ [Eq. (9)].

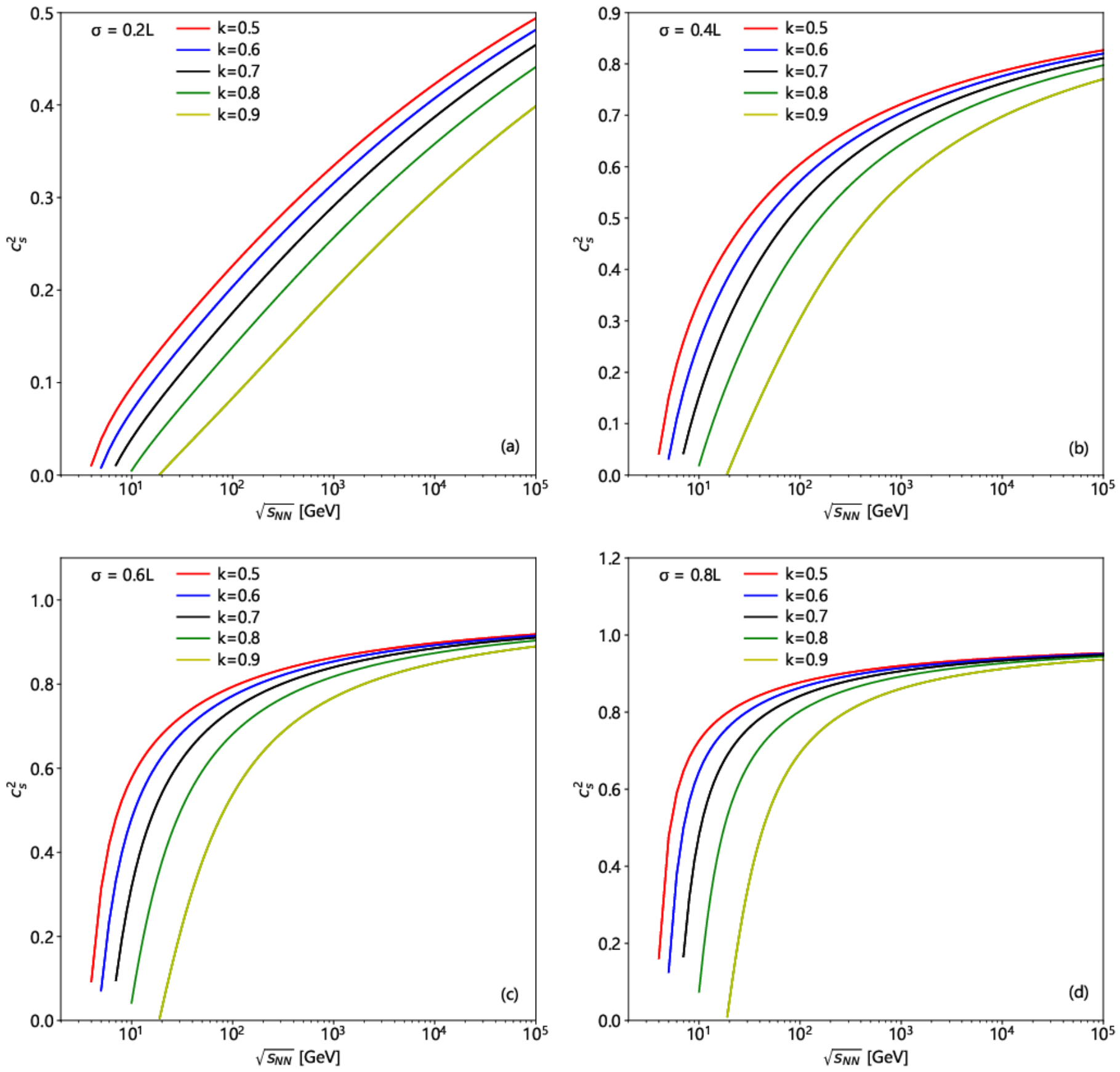

Fig. 3. Dependence of $c_s^2$ on $\sqrt{s_{NN}}$ at different $\sigma$, where $\sigma = 0.2L$ (a), $0.4L$ (b), $0.6L$ (c), and $0.8L$ (d), respectively. In each panel, $k = 0.5$, 0.6, 0.7, 0.8, and 0.9 which corresponds to various curves with different colors.

When two Gaussian distributions are required, they may be interpreted as contributions from target and projectile sources respectively. Conversely, if three distributions are necessary, these could correspondingly represent contributions from target, central, and projectile sources where ideally the contribution from the central source should dominate. This approach yields maximal values for $c_s^2$, in relation to central source while resulting in lower values for both target and projectile sources. At the energy levels currently achievable in laboratories, three Gaussian distributions are sufficient, and there appears to be no necessity for a fourth.

Regardless of the relationship between $c_s^2$ and $L$, there exist explanations regarding the magnitude of $c_s^2$. Let $d$ denote the spatial dimension; it is established that $c_s^2 = 1/d$ for matter produced in high-energy collisions [27, 28], irrespective of whether this matter exhibits liquid-like or gas-like characteristics. It should be noted that the distinction between liquid-like and gas-like states is not sharply defined; however, particles in a liquid-like state typically have a mean free path smaller than those in a gas-like state. For an ideal gas composed of massless particles with zero shear modulus in three-dimensional space, one finds that $c_s^2 = 1/3$ [29–38]. Generally speaking, for ideal gases containing massive particles, it holds that $c_s^2 < 1/3$.

In certain specific scenarios—such as when matter formed from non-central collisions resides within a thermalized cylinder aligned along the beam direction while exhibiting transverse flow without having expanded into other directions—or when considering matter situated within the reaction plane [39, 40]—it can be observed that $c_s^2 = 1/2$ for massless particles and $c_s^2 < 1/2$ for massive particles. Conversely, if matter generated from central collisions exists within such a thermalized cylinder without expansion beyond its boundaries, then one finds that $c_s^2 = 1$ for massless particles and $c_s^2 < 1$ for massive ones. Both cases where $c_s^2 < 1/2$ and where $c_s^2 < 1$ correspond to early hadron production.

As time progresses, the matter or thermalized cylinder resulting from non-central collisions expands into additional dimensions; this leads to a decrease in $c_s^2$, reducing it from below 1/2 to below 1/3. Furthermore, as the matter or thermalized cylinder originating from central collisions also expands outwardly beyond its cylindrical confines, this results in an accelerated decline of $c_s^2$ from below 1 to below 1/3. In general terms, the particle measurements obtained through experiments pertain to final states which arise after sufficient expansion has occurred, corresponding to values of $c_s^2$ that fall beneath 1/3.

However, the situation becomes more complex when strongly coupled quark-gluon plasma (sQGP) is formed in central heavy ion collisions at high energy. The formation of sQGP leads to a longer duration of the central collision system, resulting in greater expansion and a lower $c_s^2$ compared to non-central collision systems. In our view, the extraction of $c_s^2$ is highly dependent on the assumed physical models. It is crucial to adopt a reasonable and reliable physical framework along with its model representation if one aims to extract an accurate value for $c_s^2$. This work presents an alternative method for extracting $c_s^2$.

In our previous study [39], based on the Landau hydrodynamic model [11–14, 15], we investigated $c_s^2$ in both central and non-central gold-gold (Au-Au) collisions at $\sqrt{s_{NN}}$ = 19.6, 62.4, 130, and 200 GeV, as well as in copper-copper (Cu-Cu) collisions at $\sqrt{s_{NN}}$ = 22.4, 62.4, and 200 GeV by analyzing pseudorapidity distributions of charged particles. Our findings indicated that $c_s^2$ in central collisions is slightly lower than that observed in non-central collisions; however, both values are approximately within the range from 1/3 to 1/2. When examining leading nucleons within the target and projectile regions, we found that $c_s^2$ drops significantly below 1/3 due to these nucleons being in a gas-like state. These results align with those from our other prior research [40], which focused on proton-proton (pp) [proton-antiproton (p$\bar{\text{p}}$)] collisions at GeV and TeV energies. Both the works [39, 40] gave $\sigma$ and $c_s^2$ each to each, then the connection between $L$ and $c_s^2$ is established due to $L = 6\sigma$.

Before the summary, we would like to emphasize that the underlying physical reason for the relationship between $c_s^2$ and $d$ can be attributed to varying number densities of matter. In central collisions, a thermalized cylinder without additional expansion indicates a high particle number density, which should lead to a large value of $c_s^2$. However, in scenarios involving sQGP, prolonged expansion significantly reduces both density and $c_s^2$. Conversely, in non-central collisions, a thermalized cylinder exhibiting transverse flow but lacking other forms of expansion suggests an intermediate density, resulting in an intermediate value for $c_s^2$. As the collision system undergoes sufficient evolution, there is a notable decrease in density that causes $c_s^2$ to diminish to comparatively small values. The values of $c_s^2$ extracted at fixed $\sqrt{s_{NN}}$, as obtained in this work, pertain to the final stage of system evolution.

## 4. Summary

The square speeds of sound in various forms and conditions of matter within the universe have been reviewed, with a particular focus on the square speed of sound in hadronic matter that decouples from hot and dense systems formed during high-energy collisions. This study comprehensively considers factors such as energy loss of the incident beam, rapidity shifts of leading nucleons, and the application of the Landau hydrodynamic model for hadron production. Notably, the square speed of sound is correlated with the width or standard deviation of the Gaussian rapidity distribution observed in hadron production.

The extracted values for the square speed of sound range from 0 to 1/3 in most cases. For scenarios where this value exceeds 1/3 but does not surpass 1/2, we interpret these results as indicative of non-central collisions. In such cases, hot and dense matter is generated within a thermalized cylinder aligned along the beam direction while exhibiting transverse flow; however, it has yet to expand in other directions. Conversely, when values exceed 1/2 but remain below 1, we attribute these findings to central collisions wherein hot and dense matter also forms within a thermalized cylinder oriented along the beam direction without having expanded beyond its cylindrical boundaries. Over time, these concerned values rapidly decline into range from 1/3 to 1/2 and even further down into range from 0 to 1/3 due to substantial system expansion.

### Data Availability Statement

Data sharing not applicable to this article as no data sets were generated during the current study. The data used to support the findings of this study and some outcome or conclusive statements are included within the article and are cited at relevant places within the text as references.

### Ethical Approval

The authors declare that they are in compliance with ethical standards regarding the content of this paper.

## Conflicts of Interest

The authors declare that there are no conflicts of interest regarding the publication of this paper.

## Acknowledgments

The work of Shanxi Group was supported by the National Natural Science Foundation of China under Grant No. 12147215. The work of K.K.O. was supported by the Agency of Innovative Development under the Ministry of Higher Education, Science and Innovations of the Republic of Uzbekistan within the fundamental project No. F3-20200929146 on analysis of open data on heavy-ion collisions at RHIC and LHC.